\documentclass[12pt]{article}

\usepackage{geometry}                		
\usepackage{graphicx}				
\usepackage[superscript]{cite}

\title{Correlated Dynamics in Aqueous Proton Diffusion}
\author
{Sean A. Fischer,$^{\ast}$ Brett I. Dunlap, Daniel Gunlycke\\
\\
\normalsize{Chemistry Division, U. S. Naval Research Laboratory,}\\
\normalsize{Washington, DC 20375, USA}\\
\\
\normalsize{$^\ast$To whom correspondence should be addressed; E-mail:  sean.fischer@nrl.navy.mil.}
}

\date{}

\begin{document}
\maketitle

\section*{Abstract}
The aqueous proton displays an anomalously large diffusion coefficient that is up to 7 times that of similarly sized cations. There is general consensus that the proton achieves its high diffusion through the Grotthuss mechanism, whereby protons hop from one molecule to the next. A main assumption concerning the extraction of the timescale of the Grotthuss mechanism from experimental results has been that, on average, there is an equal probability for the proton to hop to any of its neighboring water molecules. Herein, we present \textit{ab initio} simulations that show this assumption is not generally valid. Specifically, we observe that there is an increased probability for the proton to revert back to its previous location. These correlations indicate that the interpretation of the experimental results need to be re-examined and suggest that the timescale of the Grotthuss mechanism is significantly shorter than was previously thought.

\section*{Introduction}

Transport of a proton through water is widely held to consist of two complementary processes: structural and vehicular diffusion. Structural diffusion occurs through the Grotthuss mechanism and gives the proton its large diffusion coefficient. The Grotthuss mechanism consists of shuttling protons through the hydrogen bond network of water. In between exchanges of the excess proton from one water molecule to the next, the total diffusion is supplemented by vehicular diffusion, which refers to the center-of-mass motion of the cation. A multitude of work has gone into understanding the details of the Grotthuss mechanism\cite{Cukierman05_876,Marx06_1848,Voth12_101,Hassanali16_7642}. Perhaps the most consistent picture of the Grotthuss mechanism is that of a generalized Eigen cation (H$_{9}$O$_{4}^{+}$) whose central hydronium ion (H$_{3}$O$^{+}$) performs a ``special pair dance" with the surrounding water molecules until a proton hop occurs and another molecule becomes the central hydronium ion\cite{Agmon08_9456}. 

While the understanding of the details of the Grotthuss mechanism has evolved over time, one aspect that has remained essentially constant is the reported timescale of the Grotthuss mechanism. Since the pioneering nuclear magnetic resonance study of Meiboom\cite{Meiboom61_375}, the timescale of the Grotthuss mechanism has widely been quoted as approximately 1.5 ps. Meiboom's NMR derived timescale was reinforced by the fact that a similar timescale is obtained when the structural component of the proton diffusion is modeled as a simple random walk\cite{Meiboom61_375,Agmon95_456}. However, this agreement should not be too surprising as the same assumptions that go into treating the Grotthuss mechanism as a simple random walk, were used by Meiboom to relate the measured nuclear spin relaxation rate to the timescale of the Grotthuss mechanism.

This is not to say that all studies of the Grotthuss mechanism have relied on the assumptions behind a simple random walk in performing analyses, indeed many have not\cite{Marx07_145901,Marx10_124108,Tuckerman09_238302,Agmon08_9456,Voth99_9361,Voth05_014506,Voth15_014104,Voth17_161719,Parrinello13_13723}. For example, Parrinello and co-workers have suggested that there could be correlation in the proton hopping directions via concerted hops along water wires\cite{Parrinello13_13723,Parrinello14_133}. If true, this would invalidate the simple random walk model for the Grotthuss mechanism; however, they did not explore the consequences of their suggestion for the interpretation of the experimental results. Additionally, we note that subsequent work by Voth and co-workers has questioned the importance of concerted proton hops\cite{Voth15_014104,Voth17_161719}. 

Of special note is a study by Halle and Karlstr\"om\cite{Karlstrom83_1047}. They employed the idea of a correlated random walk to re-examine the connection between the measured nuclear spin relaxation rate and the timescale of the Grotthuss mechanism. While they derived a model to relate the experimental relaxation rate to the hopping timescale as a function of the degree of correlation, their work was motivated by physical arguments rather than evidence of correlation and appears not to have gained favor in the literature as judged by the lack of attention their model has subsequently received. In the end, the experimental timescale for the Grotthuss mechanism has continued to be given as approximately 1.5 ps.

In the present work, we have performed \textit{ab initio} molecular dynamics simulations to address whether the simple random walk model is generally valid for the Grotthuss mechanism. In doing so, we have also provided one of the most statistically robust, \textit{ab initio} determinations of the proton diffusion coefficient to date. Our simulations clearly show correlations between proton hopping directions, suggesting that the simple random walk picture is not universally valid for the Grotthuss mechanism. Consequently, the interpretation of the experimental results for the timescale of the Grotthuss mechanism should be re-examined, with our results suggesting a substantially faster hopping time.

\section*{Methods}

For our molecular dynamics simulations our system consisted of 31 water molecules and one hydrochloric acid (HCl) molecule in a cubic box with an edge length of 9.87 {\AA}. The HCl molecule was found to dissociate rapidly and remain dissociated throughout the simulations. All calculations were performed with Quantum Espresso v5.4 using the CP module\cite{QE09}. The electronic structure was described by the PBE exchange-correlation functional in conjunction with ultrasoft pseudopotentials with 25 and 200 Ry cutoffs for the wave functions and charge density, respectively\cite{Ernzerhof96_3865,Vanderbilt90_7892,Vanderbilt91_6796,Vanderbilt93_10142}. We note that while there are well known deficiencies in the ability of the PBE functional to describe liquid water\cite{Xantheas09_221102}, previous studies have found that the underlying mechanisms of proton diffusion show only a small dependence on the choice of functional\cite{Marx07_145901,Marx10_124108,Parrinello13_13723} A Nose-Hoover chain with 4 thermostats and characteristic frequency 140 THz was used to simulate a canonical ensemble with a target temperature of either 300 or 440 K. For our Car-Parrinello molecular dynamics\cite{Parrinello85_2471}, we used a time step of 4 atomic units ($\sim$0.097 fs) and a fictitious electron mass of 300 $m_{e}$ in order to keep the propagation of the system adiabatic. Data were sampled every 10 time steps, and the first picosecond of each 8 ps trajectory was discarded for equilibration. We ran 500 independent trajectories at each temperature for a total simulation time of 8 ns. The initial configurations for each trajectory were sampled from a separate molecular dynamics simulation run with the same simulation parameters.

In order to calculate a proton diffusion coefficient, we have to define the positive charge at each point along the trajectory. There is no unique way to define molecules from a collection of atoms, and this task is even more fraught with peril for an excess charge in water\cite{Parrinello99_601}. In particular, the high frequency and amplitude of oxygen-hydrogen stretching vibrations can lead to an overabundance of molecular transitions if the definitions of molecules are too simplistic. Furthermore, since we aim to gain insight into the mechanisms of proton transport, we want to avoid potentially biasing the results through the definitions of the cation. 

The most common approach has been to identify a hydronium ion (H$_{3}$O$^{+}$) as the oxygen atom closest to three hydrogen atoms in each frame of the trajectory. Whether the positive charge is identified as the hydronium ion itself or the hydronium ion is used as a stand-in for the larger Eigen cation (H$_{9}$O$_{4}^{+}$) is often inconsequential depending on the analysis. This definition of the positive charge is susceptible to the aforementioned vibrational dynamics causing an excess of proton hops. Previous attempts to overcome this have been to simply ignore any hop that is undone by the next hop, i.e. if two successive hops result in the proton being in the same location as it was initially, those hops are ignored\cite{Marx07_145901,Marx10_124108,Tuckerman09_238302,Voth08_467,Voth17_161719}. 

This phenomenon has been referred to as proton rattling and has generally been treated as of little interest and importance. However, in the context of diffusive dynamics interpreted as a random walk, the proton hopping back to its previous site is a perfectly legitimate process. In fact, assuming a simple random walk, a third of the proton hops would be expected to undue the previous hop. Therefore, it is clear that if we hope to gain insight into dynamics of the Grotthuss mechanism, we need a cation definition that does not rely on the \textit{a priori} disregard of certain types of proton transitions in order to obtain reasonable results.

To define the protonic cation at each step, we start by assigning two hydrogen atoms to every oxygen atom based on distance. The remaining hydrogen atom, which we refer to as the excess proton, is then assigned to its closest oxygen atom. If this is the first frame of the trajectory, that hydronium ion is taken as the positive charge. If this is not the first frame, then a change of the cation only occurs if the two closest oxygen atoms to the excess proton don't include the previous hydronium oxygen atom, i.e. the excess proton is no longer in between the last hydronium oxygen atom and one of its neighbors. By limiting the hopping in this way our definition of the cation is more robust to the ``special pair dance" of the excess proton within the Eigen cation\cite{Agmon08_9456} and naturally eliminates most, if not all, false transitions due to vibrational dynamics. The hydronium ion oxygen atom was used as the location of the positive charge in the analysis.

\section*{Results and Discussion}

\begin{figure}[h]
   \centering
   \includegraphics[width=0.6\textwidth]{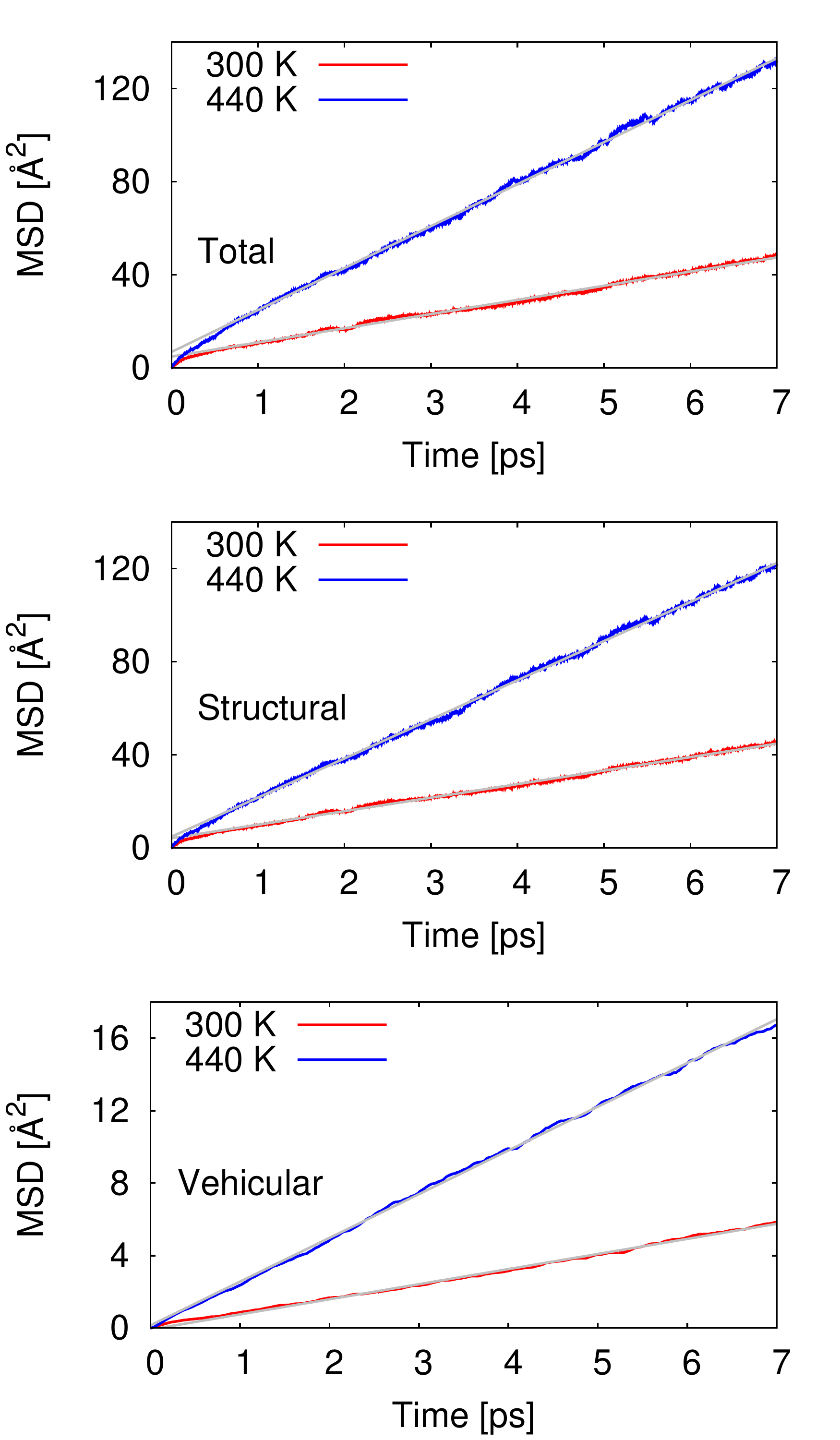}
   \caption{Mean-squared displacements (MSD) as functions of time for the proton at 300 and 440 K. The gray lines represent the linear regression used for extraction of the diffusion coefficients. The linear regression was performed on the data between 1 and 7 ps. The top panel gives the total MSD, while the middle and bottom panels show the structural and vehicular components, respectively.}
   \label{fig:msd}
\end{figure}

From our molecular dynamics simulations, we calculated the mean squared displacement of the proton as a function of time. This is shown in Fig.~\ref{fig:msd} along with the breakdown of the total into the structural and vehicular components. The diffusion coefficients were extracted from the slope of the mean squared displacement in the linear regime (between 1 and 7 ps) via the relation\cite{McQuarrie_SM} 
\begin{equation}
 D = \frac{\langle\left|\mathbf{R}(t)-\mathbf{R}_{0}\right|^{2}\rangle}{6t}
\end{equation}
where $\langle\left|\mathbf{R}(t)-\mathbf{R}_{0}\right|^{2}\rangle$ is the mean squared displacement at time $t$.
The resulting diffusion coefficients are collected in Table~\ref{tab:diff}.

\begin{table}[h]
   \centering
   \begin{tabular}{c | c c c | c c c}
      \multicolumn{1}{c}{} & \multicolumn{3}{c}{300 K} & \multicolumn{3}{c}{440 K} \\
      \hline \hline
       & D & $\sigma$ & 95\% CI & D & $\sigma$ & 95\% CI \\
      \hline
       Total & 1.015 & 0.077 & [0.860,~1.161] & 3.004 & 0.150 & [2.698,~3.291] \\
      \hline
       Structural & 0.968 & 0.070 & [0.829,~1.103] & 2.800 & 0.141 & [2.514,~3.065] \\
      \hline
       Vehicular & 0.139 & 0.007 & [0.126,~0.152] & 0.403 & 0.018 & [0.368,~0.438] \\
   \end{tabular}
   \caption{Calculated proton diffusion coefficients (D). The standard deviations ($\sigma$) and 95\% confidence intervals (95\% CI) were determined from the bootstrapping analysis using 10,000 bootstrap samples. All values are given in units of {\AA}$^{2}$/ps.}
   \label{tab:diff}
\end{table}

\begin{figure}[h]
   \centering
   \includegraphics[width=0.6\textwidth]{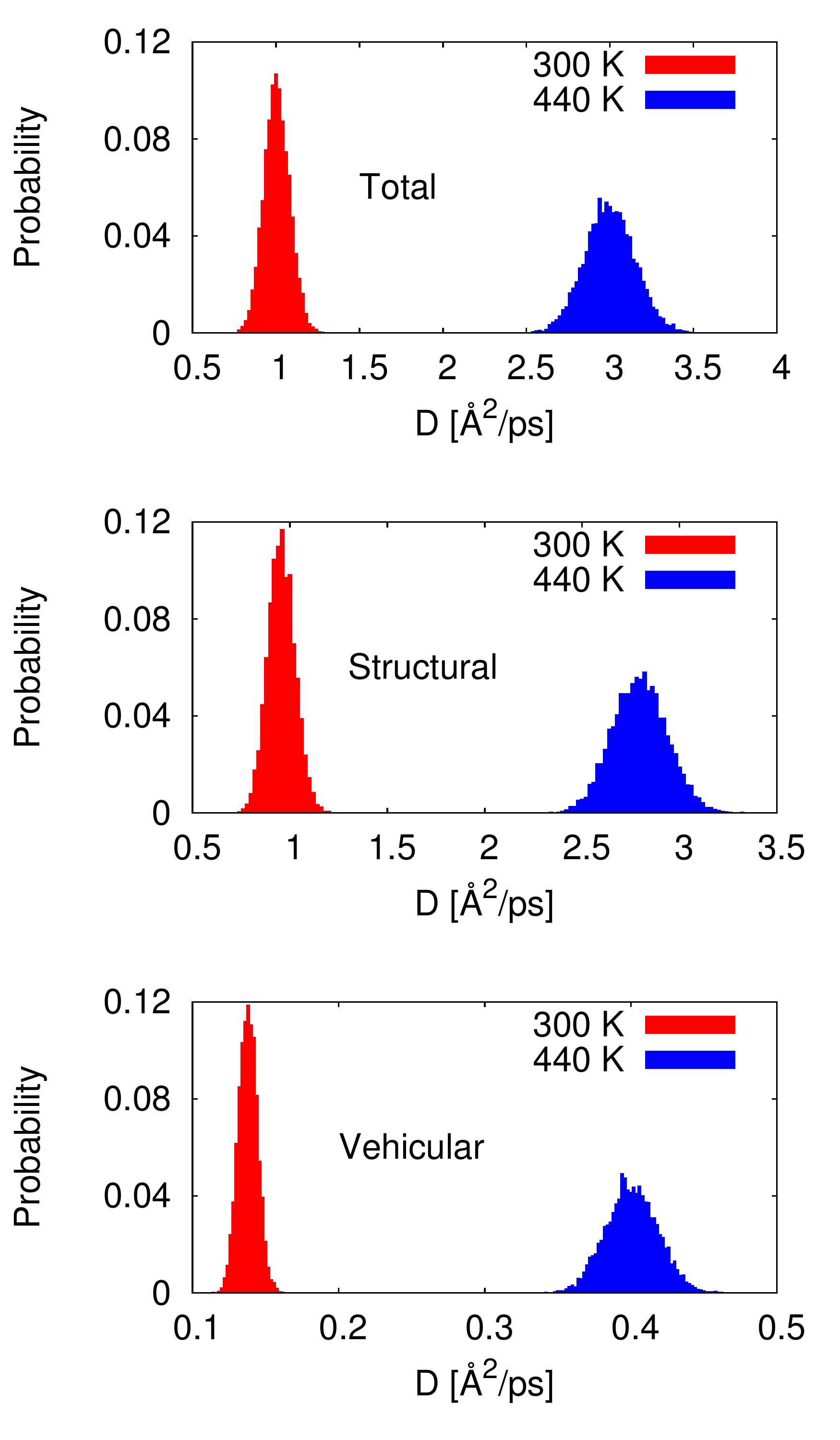}
   \caption{Distribution of the diffusion coefficients calculated in the bootstrapping analysis. Ten thousand bootstrap samples were generated, and the resulting distribution was used to calculate the standard deviations and confidence intervals given in Table~\ref{tab:diff}.}
   \label{fig:bs}
\end{figure}

Pranami and Lamm previously showed that while linear regression can be used to obtain a point estimate of the diffusion coefficient from the mean squared displacements, the statistical uncertainty of the fitting parameters are not reflective of the uncertainty in the diffusion coefficient\cite{Lamm15_4586}. To quantify the uncertainty in our calculated diffusion coefficients, we performed a bootstrapping analysis\cite{Efron79_1,Efron81_139} of the data set to obtain the standard deviations and 95\% confidence intervals that are also presented in Table~\ref{tab:diff}. The bootstrap distributions of the diffusion coefficients, from which the confidence intervals and standard deviations were derived, are shown in Fig.~\ref{fig:bs}.

\begin{figure}[h]
 \centering
 \includegraphics[width=0.7\textwidth]{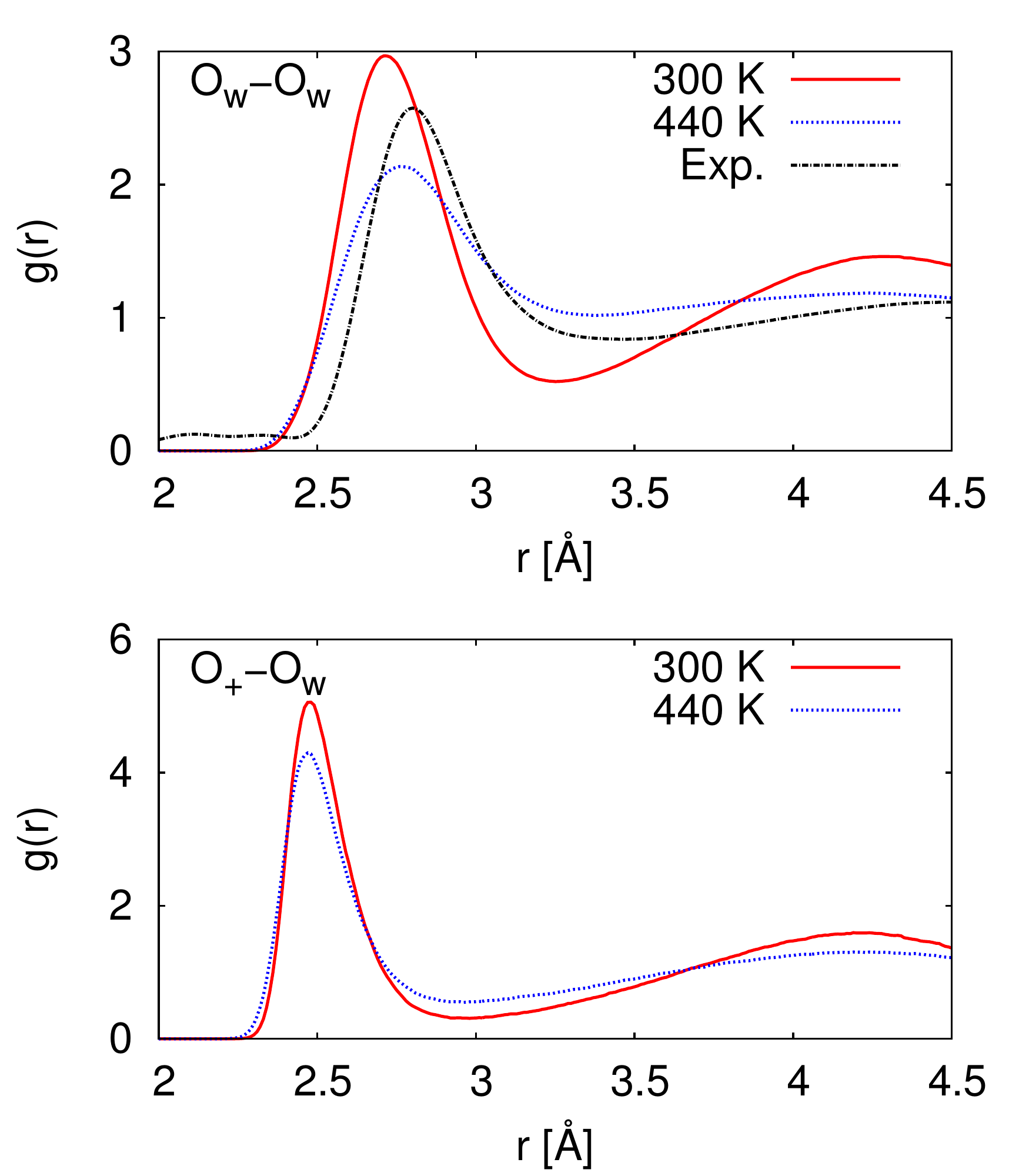}
 \caption{Radial distribution functions between water oxygen atoms (O$_{\mathrm W}$-O$_{\mathrm W}$) and between the hydronium ion oxygen atom and the water oxygen atoms (O$_{+}$-O$_{\mathrm W}$). At 300 K, the simulated water is over-structured compared to the experimental reference. While at 440 K, the simulated water is now under-structured compared to the experimental reference. The experimental reference is from Skinner et al.\cite{Benmore13_074506}, and we note that the experimental reference is for pure water while our simulations are for $\sim$1.7 M HCl.}
 \label{fig:rdf}
\end{figure}

At 300 K, our calculated proton diffusion coefficient of 1.015 {\AA}$^{2}$/ps is close to the experimental, infinite-dilution diffusion coefficient of a proton in water at ambient conditions of 0.932 {\AA}$^{2}$/ps\cite{Lobo89}. However, our simulation setup corresponds to an acid concentration of $\sim$1.7 M, and the PBE exchange-correlation functional is known to over-structure water, leading to conditions more similar to super-cooled water\cite{Xantheas09_221102}. This is evident in the oxygen-oxygen radial distribution function shown in Fig.~\ref{fig:rdf}. Under these conditions, the corresponding experimental diffusion coefficient would be smaller\cite{Sweeton41_2811,Speedy84_1888,Kreuer96_610}. That being said, our calculated proton diffusion coefficient is comparable to previously reported proton diffusion coefficients for the similarly over-structured BLYP functional (0.5 to 1.48 {\AA}$^{2}$/ps)\cite{Voth05_044505,Voth15_014104,Sebastiani17_28604}.

In order to obtain results closer to ambient conditions, we ran a second set of simulations at 440 K, which was previously suggested as a temperature at which the PBE functional gives better ambient liquid water properties\cite{Xantheas09_221102}. As can be seen in Fig.~\ref{fig:rdf}, our simulated water is now actually under-structured compared to the experimental, pure water, reference. The under-structuring is, at least partially, a result of the disruption to the water network from the excess proton and the chloride ion, as was seen in previous work on hydrochloric acid solutions\cite{Soper04_7840,Meijer06_3116}. 

As would be expected with the increase in temperature, the calculated proton diffusion coefficient is significantly larger, 3.004 {\AA}$^{2}$/ps. Again, since we are dealing with a relatively concentrated system, the corresponding experimental diffusion coefficient would still be expected to be smaller than the limiting value of 0.932 {\AA}$^{2}$/ps, by approximately a factor 1.5\cite{Sweeton41_2811}. It is safe to say that PBE overestimates the proton diffusion coefficient. 

Previous work has indicated that DFT methods underestimate the proton transfer barrier relative to wave function methods such as MP2 and CCSD(T)\cite{Voth97_7428,Singer98_5547}. By proton transfer barrier, we refer to the energetic barrier to move the excess proton from one oxygen atom to a neighboring oxygen atom. A more recent study that combined coupled cluster singles and doubles (CCSD) with path-integral molecular dynamics calculated that there was no barrier to proton transfer in the protonated water dimer\cite{Kuhne15_14355}. Additionally, nuclear quantum effects have consistently resulted in a decreased proton transfer barrier (if one existed to begin with)\cite{Voth97_7428,Parrinello99_601,Parrinello00_A153}. 

\begin{figure}[h]
 \centering
 \includegraphics[width=0.7\textwidth]{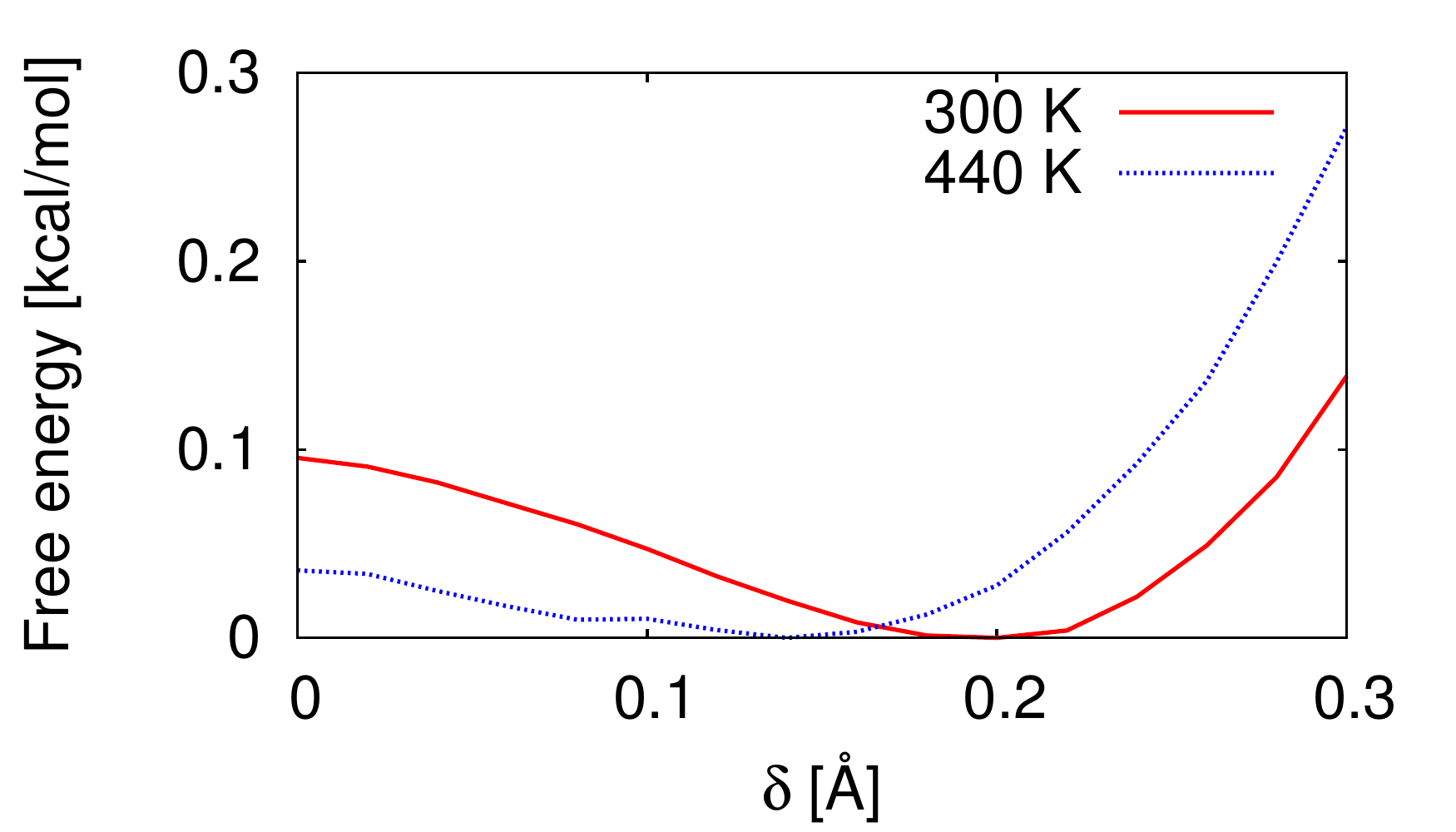}
 \caption{Calculated free energy for the proton to move from one oxygen atom to another as a function of the proton sharing coordinate ($\delta$). The proton sharing coordinate is defined as the difference between the distance from one oxygen atom to the proton and the distance between a second oxygen atom and the proton. A value of zero for this coordinate corresponds to the proton being in the middle of the two oxygen atoms.}
 \label{fig:pmf}
\end{figure}

Our calculated proton transfer barriers are displayed in Fig.~\ref{fig:pmf}. Indeed, the PBE functional gives very small barriers for the proton to transfer from one oxygen atom to another. It is possible that the small proton transfer barrier could be the origin of the overestimated diffusion coefficients; however, the proton transfer barrier is not regarded as the rate limiting step for proton diffusion: hydrogen bond dynamics to solvating water molecules are believed to control proton diffusion\cite{Agmon95_456,Parrinello95_5749,Parrinello95_150,Voth05_014506,Marx07_145901,Tuckerman09_238302} Additionally, though we only have two data points that are widely spaced, the temperature dependence of our calculated diffusion coefficients is compatible with the experimental activation energy for proton diffusion of 2-3 kcal/mol\cite{Agmon95_456} despite the proton transfer barrier being much smaller.

The calculated vehicular components to the proton diffusion coefficients are interesting in that at 300 K the vehicular diffusivity is larger than our calculated water diffusion coefficient [0.044 ($\sigma = 0.001$) {\AA}$^{2}$/ps], yet at 440 K the vehicular component is smaller than the calculated water diffusion coefficient [0.562 ($\sigma = 0.007$) {\AA}$^{2}$/ps]. The most likely explanation is that at 300 K when the water is over-structured, the water molecules are hindered in their diffusion. At the same time, the ``special pair dance" of the proton causes an elevated diffusion as the central hydronium ion rattles around within its first solvation shell\cite{Agmon08_9456}. When the temperature is elevated and the water molecules are more dynamic, the contribution of the ``special pair dance" is not as prominent.

A common assumption concerning the proton diffusion coefficient has been that the structural and vehicular components are independent, i.e. structural plus vehicular equals total. Our current results suggest that this is not the case. While the differences between the sums of the components and the totals are small (0.092 at 300K and 0.199 at 440K), our bootstrapping analysis indicates that these differences are statistically significant at the 95\% confidence level. This type of correlation between the components of the diffusion process has been suggested before based on physical arguments surrounding the polarization resulting from the hopping of the charge from one site to another\cite{Karlstrom83_1047}. In that study, the correlation was estimated to be of the order of the vehicular component of diffusion, in agreement with our current simulations.

\begin{figure}[h]
 \centering
 \includegraphics[width=0.7\textwidth]{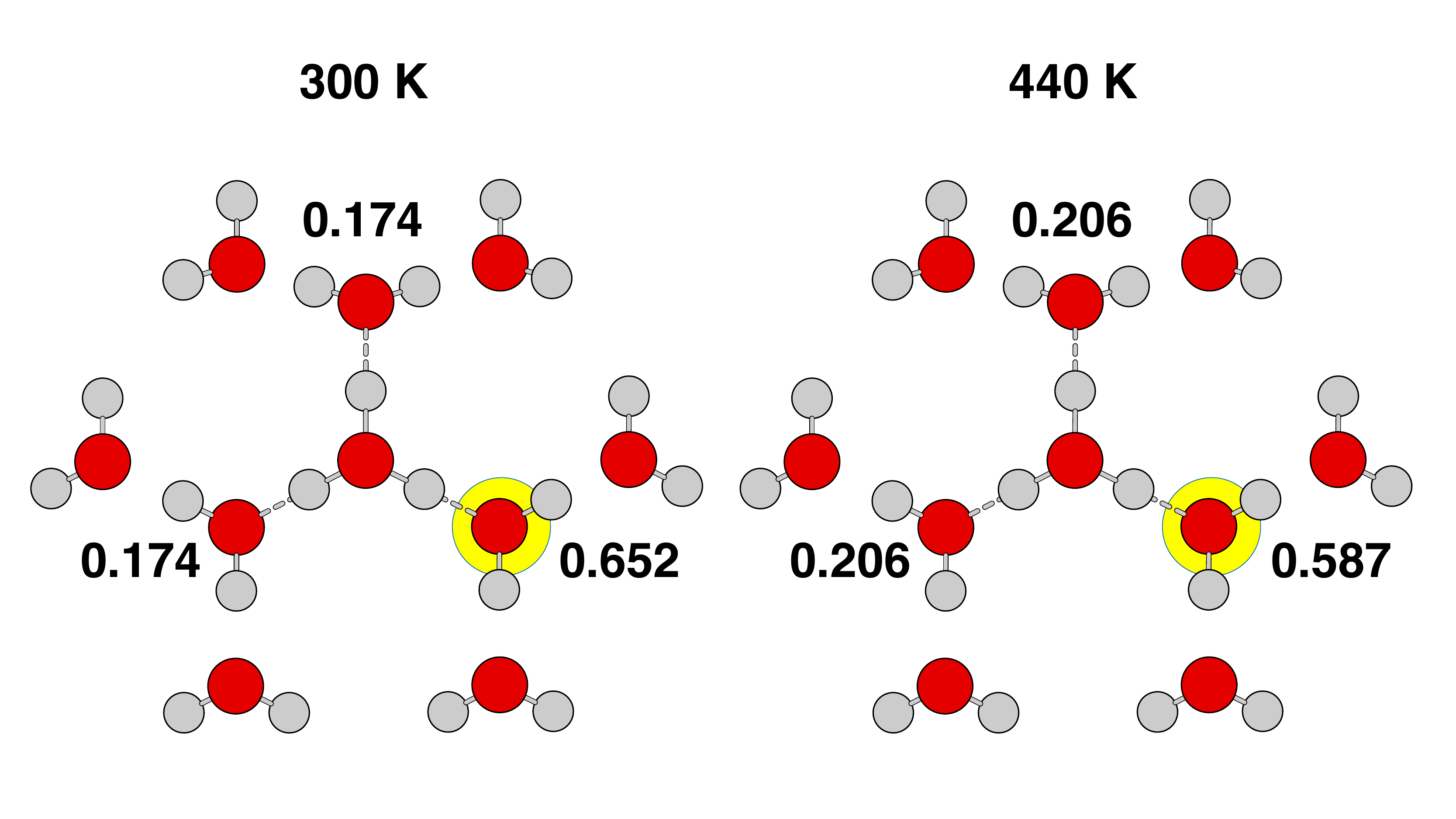}
 \caption{Illustration of the structure of the Eigen cation, along with an indication of the correlations observed in the dynamics. The atom highlighted in yellow represents the central atom of the previous cation. The numbers are the probabilities for the direction of the next proton transition. Standard deviations for the reported probabilities are given in the text.}
 \label{fig:hop}
\end{figure}

While correlation between the components of the diffusion process is noteworthy, it does not have any bearing on the validity of the simple random walk assumption for interpretation of the experimental results. For the simple random walk picture to be valid for the Grotthuss mechanism, the probability for the proton to hop to any of its three neighbors should be equal and not depend on the proton's history. Figure~\ref{fig:hop} shows the observed probabilities for the proton to hop to its neighboring water molecules, given that before the previous hop it was located on the molecule indicated by the yellow circle. Our simulations clearly reveal that there is a strong preference to return to the previous site as opposed to continuing on to a new site. Though decreased slightly at the elevated temperature, the correlation is robust, suggesting that a simple random walk is not an adequate model for the Grotthuss mechanism. 

Note that the correlation in the proton hopping directions we observe here is the opposite of that suggested by Parrinello and co-workers\cite{Parrinello13_13723,Parrinello14_133}. While we can find examples in our set of trajectories where the proton bursts across multiple water molecules in a short timeframe, the overall statistics clearly show that there is an enhanced probability for the proton to revert back to its previous location at any given step in the dynamics. This illustrates the importance of sufficient sampling as individual trajectories can give a misleading picture and obscure the underlying dynamics.

The implications of needing to go beyond the simple random walk model for interpreting the experimental results can be substantial. The details of the relationship between a simple random walk and the corresponding correlated random walk are known\cite{Renshaw94_869}. The most relevant result is that the mean squared displacement of a correlated random walk to related to the mean squared displacement of the simple random walk by a ratio of probabilities
\begin{equation}
 \langle\Delta\mathbf{R}^{2}\rangle_{c} = \left(\frac{1+p-q}{1+q-p}\right)\langle\Delta\mathbf{R}^{2}\rangle_{s}
\end{equation} 
where $\langle\Delta\mathbf{R}^{2}\rangle_{c}$ represents the mean squared displacement of the correlated walk, $\langle\Delta\mathbf{R}^{2}\rangle_{s}$ represents the mean squared displacement of the simple walk, $q$ is the probability to reverse the previous hop, and $p$ is the probability to hop to one of the other sites. Assuming an average hop length $\varepsilon$ and hopping time $\tau$, we can relate the diffusion coefficient to these quantities using the general random walk model
\begin{equation}
 D = \left(\frac{1+p-q}{1+q-p}\right)\frac{\varepsilon^{2}}{6\tau}
\end{equation}
where in the case of a simple random walk $p=q$.

As indicated in Fig.~\ref{fig:hop}, our simulations give $p=0.174~(\sigma=0.048)$ and $q=0.652~(\sigma=0.096)$ at 300 K; $p=0.206~(\sigma=0.030)$ and $q=0.587~(\sigma=0.061)$ at 440 K. Using the equilibrium distance between the hydronium ion and a water molecule of 2.5 {\AA} (see Fig.~\ref{fig:rdf}), along with our calculated structural diffusion coefficients from Table~\ref{tab:diff}, the correlated random walk model gives a hopping timescale of 0.380 ps at 300 K and 0.167 ps at 440 K. Using our calculated hopping probabilities with the experimental diffusion coefficient of 0.932 {\AA}$^{2}$/ps, we get a hopping timescale of between 0.460 and 0.665 ps, depending on the assumed vehicular component (sodium ion, 0.133 {\AA}$^{2}$/ps; water molecule, 0.230 {\AA}$^{2}$/ps\cite{Lobo89}) and which set of probabilities is used. Clearly this range is substantially different from the 1.304 to 1.484 ps that is obtained from assuming a simple random walk. Note also that here we have not taken into account the correlation between the components that we found in our simulation. Doing so could lead to even faster timescales as the correlation implies that the structural component could be even larger than what has been assumed for the experimental case.

\begin{figure}[h]
 \centering
 \includegraphics[width=0.7\textwidth]{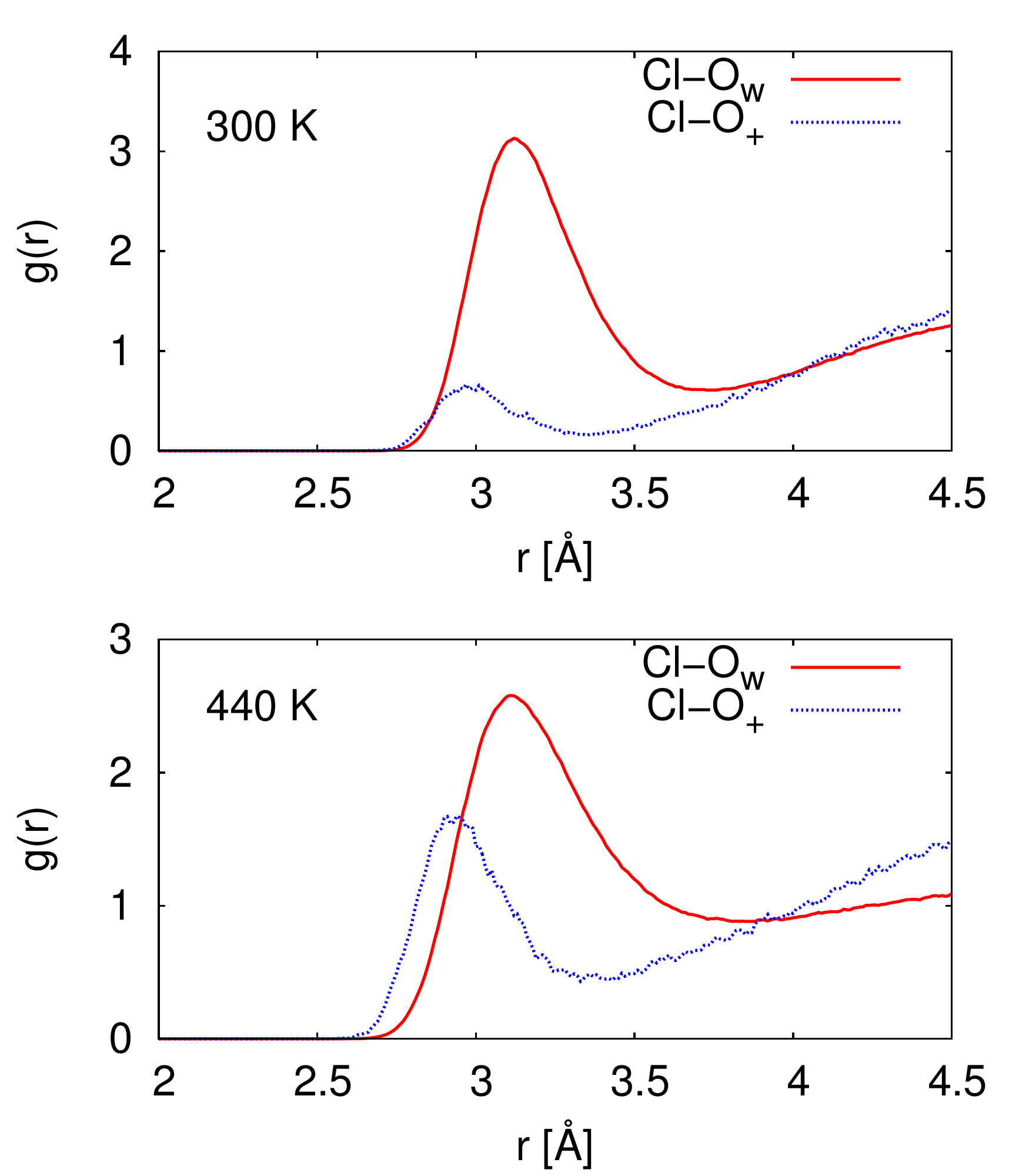}
 \caption{Radial distribution functions between the chloride ion and water oxygen atoms (Cl-O$_{\mathrm W}$) and between the chloride ion and the hydronium ion oxygen atom (Cl-O$_{+}$). The peak around 3 {\AA} is due to the first solvation shell of the chloride ion, indicating some direct interaction with the hydronium ion.}
 \label{fig:clrdf}
\end{figure}

\begin{figure}[h]
 \centering
 \includegraphics[width=0.7\textwidth]{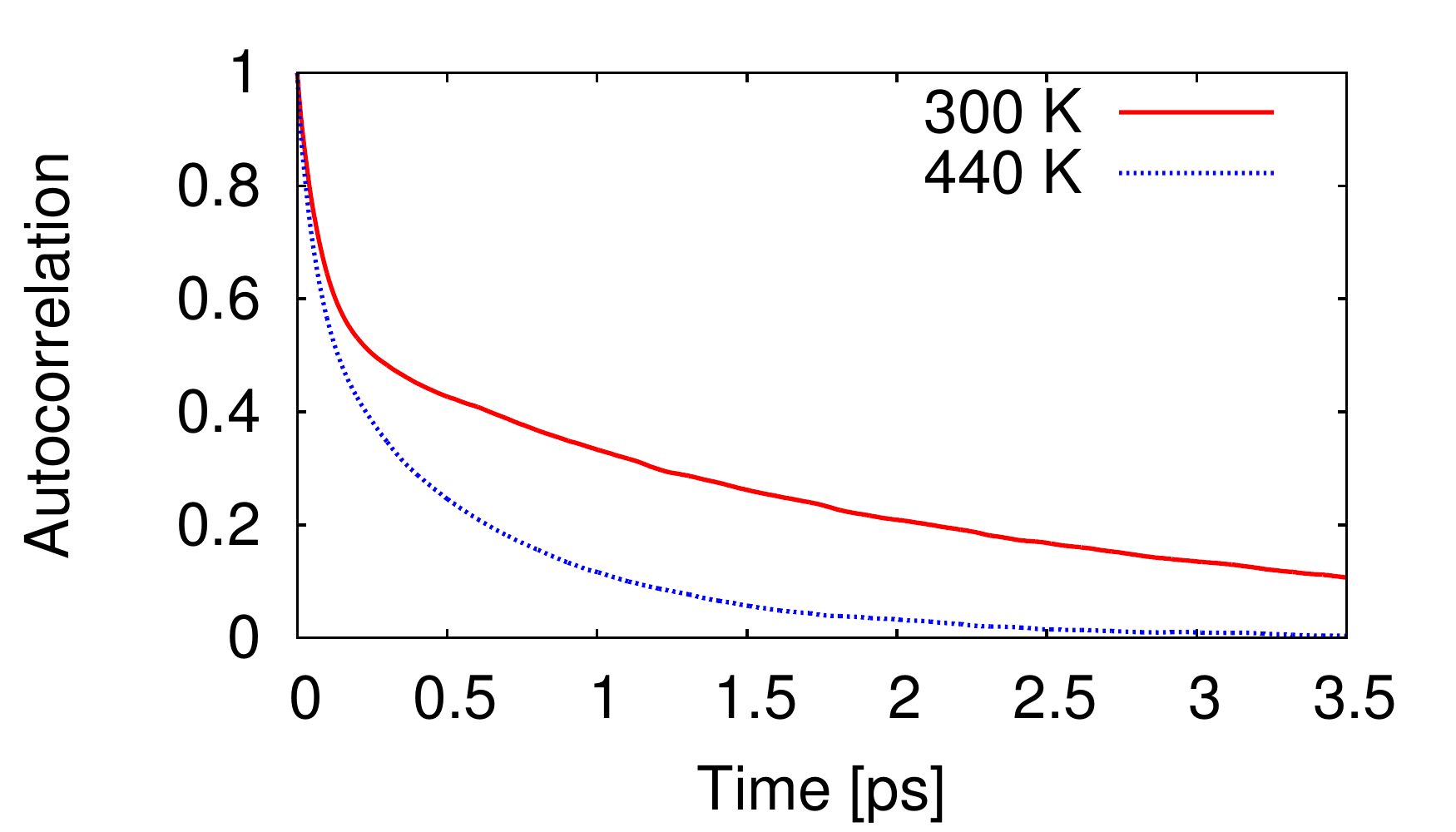}
 \caption{Normalized autocorrelation for the vector connecting the hydronium and chloride ions. Though the hydronium and chloride ions form contact pairs more often at 440 K (see Fig.~\ref{fig:clrdf}), the correlation between the two ions decays more quickly at higher temperature.}
 \label{fig:clcor}
\end{figure}

Given that our simulations were performed at an HCl concentration of $\sim$1.7 M, it is reasonable to wonder whether the correlations we observe can be applied to the infinite-dilution case. While a more complete answer to that question would require additional investigations, we note that the concentration of our system sits right on the edge of where changes begin to occur in the experimental vibrational spectra\cite{Ben-Amotz17_5246}. In our simulations we do find that the chloride ion and hydronium ion occasionally form contact ion pairs (Fig.~\ref{fig:clrdf}), as has been seen before\cite{Meijer06_3116}, and we observe some relatively long-lived correlations in the vector connecting the hydronium and chloride ions (Fig.~\ref{fig:clcor}). As such, we cannot rule out ion-ion interactions leading to some of the correlation that we observe in the hopping directions. However, we note that spectroscopic studies aiming to gain insight into aqueous proton dynamics have been done at even higher concentrations than we have studied here\cite{Bakker06_138305,Havenith15_11898,Tokmakoff15_78,Tokmakoff17_154507}, providing obvious relevance for having a reliable model for the dynamics of protons in more concentrated solutions.

\section*{Conclusions}

Through large sets of \textit{ab initio} molecular dynamics simulations, we have found significant correlation between hopping directions in the Grotthuss mechanism of aqueous proton diffusion. Specifically, we found an elevated probability for the proton to return to its previous site compared to what would be expected for a simple random walk. These results suggest that the interpretation of the experimental results for proton diffusion needs to be re-examined. Until now, the experimental results have generally been interpreted in terms of a simple random walk, resulting in a timescale of approximately 1.5 ps for the Grotthuss mechanism. However, re-interpreting those results in terms of the correlated random walk suggested by our simulations, results in the timescale being closer to 0.5 ps. Furthermore, our results also provide evidence of correlation between the components of the diffusion coefficient. This could mean that the timescale of the Grotthuss mechanism is even faster since we found a negative correlation, meaning that the individual components add to more than the total. While we have found that the correlations between the components of the diffusion and the hopping directions are robust to temperature, further work is needed to assess the dependence of these correlations on concentration.

\section*{Acknowledgements}
The authors acknowledge support from the U.S. Office of Naval Research through the U.S. Naval Research Laboratory.




\subsection*{Competing financial interests}
The authors declare no competing financial interests


\end{document}